\documentstyle[eqsecnum,aps,prb,floats,epsfig]{revtex}
\begin{document}
\twocolumn

\title{Fragility of the spin-glass-like collective state to a
magnetic field in an interacting Fe-C nanoparticle system}

\author{P. E. J{\"o}nsson, S. Felton, P. Svedlindh, and P. Nordblad}
\address{
Department of Materials Science, Uppsala University\\
        Box 534, SE-751 21 Uppsala, Sweden}

\author{M. F. Hansen}
\address{Department of Physics, Building 307, Technical University of
Denmark\\  DK-2800 Lyngby, Denmark}

\date{\today}
\maketitle

\begin{abstract}
The effect of applied magnetic fields on the collective
nonequilibrium dynamics of a
strongly interacting Fe-C nanoparticle system has been investigated.
It is experimentally shown that the magnetic aging diminishes to
finally disappear for fields of moderate strength.
The field needed to remove the observable aging behavior increases with
decreasing temperature. The same qualitative behavior is observed in
an amorphous metallic spin glass
(Fe$_{0.15}$Ni$_{0.85}$)$_{75}$P$_{16}$B$_6$Al$_3$.

\end{abstract}


Frozen ferrofluids offer systems where the long-range dipolar
interaction between the single-domain nanoparticles can be
continuously varied by changing the particle concentration.
In a dilute ferrofluid, where the dipolar interaction energy is
negligible compared to the anisotropy energy, the magnetic properties
of the system is given by averaging over the individual particle
contributions.
The dynamical properties of isolated particles are determined by the
thermally activated relaxation between the potential wells of the
anisotropy energy, as was originally proposed by N{\'e}el.\cite{nee49}
The magnetic response of a single-domain particle is  strongly affected
by an external bias field. 
The relaxation time depends both on the magnitude and the direction
of the applied magnetic field with respect to the anisotropy axis, in
combination with the damping of the gyration around the easy axis after a spin-flip (see, e.g., Refs.\ \onlinecite{bro63,aha69,garetal99,cofetal2000}).
For interacting particle systems, the dipolar field created by surrounding particles will also affect the relaxation time. \cite{jongar2001EPL}

Highly concentrated ferrofluids, in which the dipolar interaction
energy dominates over the anisotropy energy, contain all
ingredients needed to create collective glassy dynamics; a complex
interaction mechanism - the dipolar interaction - and frustration
provided by the randomness of the particle positions and directions
of the anisotropy axes.
Indeed, experiments have shown that such particle systems exhibit
nonequilibrium dynamics with striking similarities to the
nonequilibrium dynamics of spin glasses. The magnetic relaxation of
the low field dc magnetization shows an aging
behavior\cite{jonetal95} and there is a downward relaxation of the 
low-frequency ac susceptibility when the sample is kept at constant
temperature combined with a rejuvenation and memory behavior when
the temperature of the sample is further decreased and
subsequently re-heated.\cite{mametal99,jonetal2000PRB}
Also, certain effects of applied magnetic fields on the nonequilibrium
behavior of particle systems have been reported. 
In a recent work it was shown that the relaxation of the
thermoremanent magnetization (TRM) exhibits magnetic
aging if the field applied during cooling is low, but not if a relatively
high field is used. \cite{garmuretal99} 
This result is similar to the behavior of spin glasses where large
enough fields impose an equilibrium state on experimental time scales
and a relaxation of the thermoremanent magnetization that is independent of wait
time \cite{graetal87} and it is thus also consistent
with the assumption of collective
spin-glass-like dynamics in strongly interacting nanoparticle
systems. \cite{jonnor2000}
In the spin glass case, it
has also been shown that field changes can induce rejuvenation of the spin glass state \cite{djuetal94} and that the nonequilibrium relaxation is affected by large enough fields.\cite{djuetal95,johetal96}

In this article, we study the effect of bias magnetic fields on the
nonequilibrium dynamics of a strongly interacting nanoparticle sample.
We have chosen to measure the relaxation of the low-frequency ac
susceptibility, since such relaxation is indicative of magnetic aging
and therefore does not exist in weakly interacting nanoparticle
systems.
It is observed that the collective nonequilibrium dynamics disappears at
moderate
fields and that the strength of the field needed to remove the
collective glassy dynamics increases with decreasing temperature.
The same qualitative behavior is observed for a spin glass sample.
We interpret the results within the droplet model, \cite{fishus88}
which is a real-space model that has successfully been used to
describe nonequilibrium effects in spin glasses.

The ferrofluid consisted of ferromagnetic nanoparticles of the
amorphous
alloy Fe$_{1-x}$C$_x$ ($x\approx0.2-0.3$).
A TEM study revealed a nearly spherical particle shape and a particle
size of $d=5.3\pm0.3$~nm. The saturation magnetization is estimated to $M_s=1 \cdot 10^3$~G and the anisotropy constant to $K=9\cdot 10^5$~erg\,cm$^{-3}$. 
Details about the sample preparation and characterization are given elsewhere. \cite{hanetal2001}
Two samples were studied, one with a concentration of 5~vol\%
which has earlier been shown to exhibit spin-glass-like nonequilibrium
dynamics, \cite{jonetal2000PRB} while the second sample was diluted
to 1~vol\% and no magnetic aging was observed in this sample.

For comparison, experiments were also performed on an amorphous metallic
spin glass
(Fe$_{0.15}$Ni$_{0.85}$)$_{75}$P$_{16}$B$_6$Al$_3$. 
The interaction in this sample is of RKKY type and it behaves as a dilute
metallic spin glass alloy with a transition temperature of
$T_g=22.5$~K.
Various magnetic properties of the system have been investigated in a
number of earlier reports (see, e.g., Refs.
\onlinecite{sveetal86,sveetal87,djuetal94,djuetal95}).
We chose this compound since it has a high
susceptibility and comparably low applied magnetic fields are needed
to affect the nonequilibrium dynamics.

The temperature dependence of the ac susceptibility of the two nanoparticle
samples with a superimposed dc field was measured in a
LakeShore 7225 ac susceptometer using an ac probing field of
frequency 125~Hz and amplitude 2~Oe and dc fields in the range 0 --
250 Oe. These fields are low enough not to destroy the two-well
structure of the single-particle potential.\cite{Hc}
A noncommercial low-field SQUID magnetometer \cite{magetal97} was
used to measure the ac susceptibility with superimposed dc fields as
a function of time on both the 5 vol\% nanoparticle sample and the
spin glass. 
The samples were cooled in the dc field from a temperature
in the paramagnetic region to the measuring temperature, and the data collection was initiated after waiting a short time $t_0$ allowing the system to stabilize.
For the nano-particle sample $t_0=120$~s and for the spin glass $t_0=60$~s.
The frequency of the ac field was 510~mHz, the field amplitude 10~mOe and the range of dc fields 0-240~Oe.

The total energy of a nanoparticle system, probed by an ac field of amplitude
$h_0$ and with an applied dc field $H$ in the same direction as the ac
field, is given by
\begin{equation}
E = E_{\rm a} -\sum_i\vec{m}_i \cdot (h_{0} \hat{z} \sin \omega t +
H \hat{z} + \vec{H}_{i}^{\rm dip}) \; ,
\end{equation}
where $E_a$ is the anisotropy energy and $\vec{m}_i$ is the
magnetic moment of particle $i$.
The dipolar field at the position of the $i$th particle is given by
\begin{equation}
\vec{H}_{i}^{\rm dip} = \frac{1}{4 \pi}
\sum_{j} \left[ \frac{3(\vec{m}_j \cdot
\vec{r}_{ij})\vec{r}_{ij}}{r_{ij}^5} -\frac{\vec{m}_j}{r_{ij}^3}
\right] \; ,
\end{equation}
where $j \ne i$ and $\vec{r}_{ij}=\vec{r}_i-\vec{r}_j$ is the vector
connecting particle $i$ with particle $j$.

\begin{figure}[htb]
\centerline{\epsfig{figure=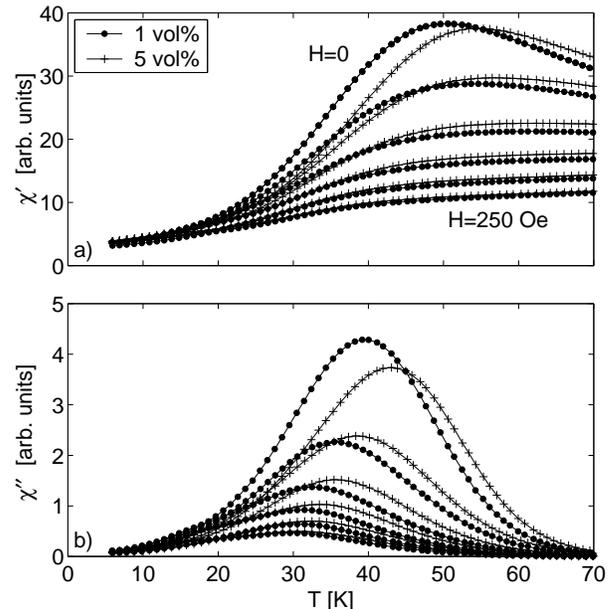,width=8cm}}
\caption[]{The ac susceptibility vs. temperature for different
superimposed dc fields; H = 0, 50, 100, 150, 200, 250 Oe. $f=125$~Hz.}
\label{suscLS}
\end{figure}

The results from measurements of the ac susceptibility with superimposed
dc fields for the two nanoparticle samples are
shown in Fig.~\ref{suscLS}.
For $H = 0$ the peak of the ac susceptibility appears at higher
temperature for the 5 vol\% sample than for the 1 vol\% sample, due
to the stronger dipolar interactions in the more concentrated sample.
However, the difference between the susceptibility of the two samples
decreases with increasing dc field, and for the highest fields the
susceptibility curves are almost identical indicating that the dipolar field created by the surrounding particles is negligible compared to the applied field.
Corresponding ac susceptibility results with
bias fields for the spin glass sample are shown in Fig. 1 of
Ref. \onlinecite{sveetal87}.

\begin{figure}[htb]
\centerline{\epsfig{figure=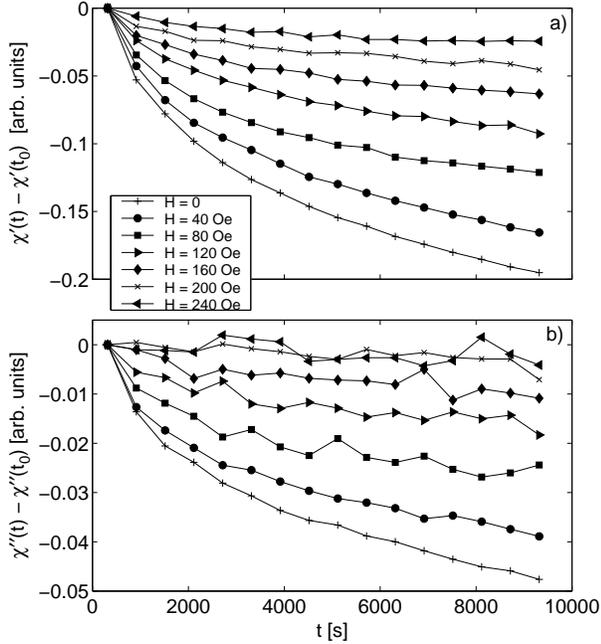,width=8cm}}
\caption[]{Relaxation of the ac susceptibility
for different superimposed dc fields $H$,
measured on the 5vol\% nanoparticle sample. The frequency
$\omega/2\pi=510$~mHz and the temperature $T=25$~K. Same units as in
Fig.~\ref{suscLS}.}
\label{ac-relax}
\end{figure}

We have chosen to study how the collective behavior is affected by a
magnetic field by measuring the relaxation of the low-frequency ac susceptibility in superimposed dc fields.
Aging effects are seen in $\chi''(\omega)$ as a
relaxation towards equilibrium with time spent at constant
temperature, $t$ ($\gg 1/\omega$).\cite{lunetal81}
The corresponding relaxation is seen in $\chi'(\omega)$, and since the relaxation is
larger in $\chi'$ than in $\chi''$ it can be more
convenient to study $\chi'$ if the relaxation is
small.\cite{jonetal2000PRB}
Fig.~\ref{ac-relax} shows $\chi(H,t)-\chi(H,t=t_0)$ for different dc
fields at $T=25$~K for the 5 vol\% sample.
It is seen that the field reduces the relaxation and at fields higher
than 200~Oe there is almost no relaxation left.
A reduction of the relaxation in the ac susceptibility with applied
magnetic fields is also observed for the spin glass sample.

\begin{figure}[htb]
\centerline{\epsfig{figure=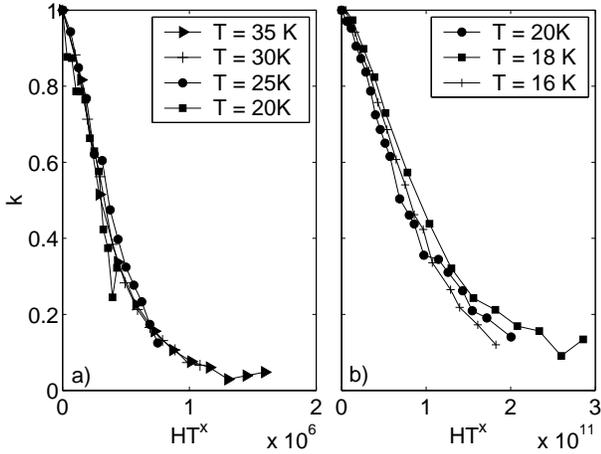,width=8cm}}
\caption[]{The quantity $k(HT^{x})$ at different temperatures for a) the 5 vol\% nanoparticle sample using $x=2.5$ and b) the spin glass using $x=7.5$.
}
\label{Fig: k}
\end{figure}

We now define a quantity $k$ as
\begin{equation}
k(H) = \frac{\chi(H,t=t_0)-\chi(H,t=t_{\rm max})}
	{\chi(H=0,t=t_0)-\chi(H=0,t=t_{\rm max})} \; ,
\label{Eq: k}
\end{equation}
which gives a relative measure of the relaxation in the presence
of a dc field.
We have repeated the measurements in Fig.~\ref{ac-relax} at
different temperatures for both the 5 vol\% sample and the spin glass
sample. For both samples the relaxation persists to higher fields at
lower temperatures.
Postulating an $HT^{x}$ dependence of $k$ and using $x$ as a fitting
parameter we obtain reasonable scaling behavior for the two samples.
In Fig.~\ref{Fig: k}(a) $k(H T^{2.5})$ is shown for temperatures in
the interval 20-35 K, for the 5 vol\% nanoparticle sample. 
The curves measured at different temperatures give a satisfactory
data collapse.
For the spin glass sample, on the other hand, data
collapse is obtained for
curves measured at different temperatures between 16 and 20 K using
$k(H T^{7.5})$, as can be seen in Fig. \ref{Fig: k}(b). 
The choice of using $\chi'$ or $\chi''$ in the analysis did not affect $k$.

We will interpret our results within the droplet model
\cite{fishus88} which was derived for short-range Ising spin glasses.
Important concepts of the model should however also be applicable to particle
systems exhibiting strong dipole-dipole interaction and random
orientation of the anisotropy axes.
In this model, it is assumed that, at each temperature below the spin-glass transition temperature $T_g$, the spin-glass equilibrium state is unique but two-fold degenerate by its global spin-reversal state.
In equilibrium, the most important contribution to physical observables, such as the magnetic susceptibility, comes from low-lying excitations of compact clusters of spins, called {\it droplets}.
The droplet excitations of size $L$ have a broad distribution of free energies, with a typical value of
\[
F_L^{\rm typ} \sim \Upsilon (L/L_0)^\theta \, ,
\]
where $\Upsilon(T)$ is the stiffness modulus, $\theta$ is the stiffness exponent, and $L_0$ is a characteristic length scale. 
The stiffness exponent satisfies $\theta \leq (d-1)/2$, where $d$ is the dimension of the system, and $\theta \approx 0.2$ for $d=3$. \cite{bramoo84,kometal99}
The dynamics of droplets is a thermally activated process.   
The typical energy barrier scales as
\[
B_L^{typ} \sim \Delta (L/L_0)^\psi \, ,
\]
where $\Delta(T)$ sets the free energy scale of the barriers and the barrier exponent $\psi$ satisfies $\theta \le \psi \le d-1$. The value of the exponent $\psi$ has been estimated to 0.8.\cite{matetal95}

Let us now consider the isothermal aging process which we experimentally have observed as a relaxation of the low-frequency ac susceptibility.
Within the droplet model, the development towards equilibrium from the out-of-equilibrium state, which was created when quenching the system, is governed by the growth of domains belonging to either of the two degenerate equilibrium states.
That growth is driven by successive nucleation and annihilation of droplets. 
The growth law proposed by Fisher and Huse\cite{fishus88} is
\begin{equation}
\label{L-FH}
L_T(t) \sim L_0 \left[ \frac{T \ln(t/\tau_0)}{\Delta(T)}\right]^{1/\psi} \,,
\end{equation}
where $\tau_0$ is the relaxation time of a spin (or magnetic moment).
The weak ac field of frequency $\omega/ 2 \pi$ probes the system by polarizing droplets of size $L_T(1/\omega)$.
Since $t \gg 1/\omega$, $L_T(1/\omega)<L_T(t)$ and hence the domain walls of size $L_T(t)$ appear effectively frozen on the probing length scale.
The small scale droplets [of size $L_T(1/\omega)$] in touch with a frozen-in domain wall will have a lower free energy than they would have had if the frozen-in wall was not present. This can be described by a reduction of the effective stiffness of the system. Fisher and Huse used scaling arguments to obtain 
\[
\frac{\Delta \Upsilon}{\Upsilon} \sim 
\left[
\frac{L_T(1/\omega)}{L_T(t)}
\right]^{d-\theta}
\]
and deduced that
\begin{equation}
\label{Xrelax}
\chi''(\omega) = \chi_{eq}''(\omega)
\left\{
1-c \left[
\frac{L_T(1/\omega)}{L_T(t)}
\right]^{d-\theta}
\right\} \, ,
\end{equation}
where $c$ is a constant.
It has been shown experimentally, for a 2d Ising spin glass, that both $\chi''$ and $\chi'$ relax according to this expression. \cite{Schetal93}

In a magnetic field $H$, the system is disordered by the field
on length scales larger than the correlation length 
\[
\xi_H \sim
\left[\frac{\Upsilon}{H \sqrt{q_m}}\right]^{2/(d-2\theta)} \,,
\]
 while it still exhibits spin glass order on shorter
length scales. Here, $q_m(T)$ is an order parameter defined in Ref.~\onlinecite{fishus88}.
The typical time needed for the system to equilibrate is given by $t_{eq} \sim \tau_{H}$,
where $\ln (\tau_H/\tau_0) \sim \frac{\Delta}{T} \xi_H^\psi$. 
The relaxation of $\chi$ at a certain temperature will then be governed by
the relation between the domain size $L_T(t)$ reached within
the experimental time window and the length scale $\xi_H$ set by the
magnetic field. 
In a strong applied field Eq.~(\ref{Xrelax}) can be modified to include the correlation length, as
\begin{equation}
\label{Xrelax-H}
\chi''(\omega,H) = \chi_{eq}''(\omega,H)
\left\{
1-c \left[
\frac{L_T(1/\omega)}{min(L_T(t),\xi_H)}
\right]^{d-\theta}
\right\} \, .
\end{equation} 
Here, it should be noted that neither $L_T(t)$ nor $\xi_H$ are well defined
length scales, so the relaxation will not end abruptly, but will gradually be suppressed over a wide time window.
Three different field regimes can be distinguished:
i) $L_T(t) \ll \xi_H$, the collective nonequilibrium dynamics is virtually 
unaffected by the field.
ii) $L_T(t) \lesssim \xi_H$, the system is partly at equilibrium, and
hence the ac relaxation is reduced.
iii)  $L_T(t) > \xi_H$, the system is in equilibrium, no collective
dynamics exists.

At sufficiently low temperatures ($T \lesssim T_g/2$) the influence of critical fluctuations is small and hence the temperature dependent coefficients $\Upsilon$, $\Delta$, and $q_m$ are approximately constant. \cite{fishus88}
Since $k(H)$ scales with $L_T(t)/\xi_H$ according to Eq.~(\ref{Xrelax-H}), we then expect to obtain  data collapse, at low temperatures, plotting $k$ vs $HT^{\tilde{x}}$, with $\tilde{x}=(d-2\theta)/2\psi$.
The reported values of $\theta$ and $\psi$ yield $\tilde{x}>1$ and hence the condition $L_T(t)/\xi_H \approx 1$ is fulfilled for lower fields at higher temperatures in accordance with results shown in Fig.~\ref{Fig: k}.
Due to experimental limitations, the measurements on the spin glass correspond to $T>T_g/2$ and therefore $x \ne \tilde{x}$, since critical fluctuations will modify the temperature dependence.

The quantitative difference, between the spin glass and the nanoparticle
system, in how the field needed to affect the collective dynamics
scales with temperature can be attributed to the difference in
relaxation time of isolated spins in the two systems.
The spin flip time of an atomic moment in the spin glass is constant,
while for the magnetic moment of a nanoparticle the relaxation time
depends both on the temperature and the field in a nontrivial way, and is also affected by the dipolar interactions.

We have shown, by measuring the isothermal relaxation of the ac
susceptibility with superimposed dc fields, that the collective
glassy dynamics of a strongly interacting nanoparticle system can be
destroyed by the application of moderate fields. The field strength
needed to destroy the collective dynamics increases with decreasing
temperature. This behavior is consistent with corresponding observations on a
spin glass sample. The results for both samples show quantitative agreement with predictions within the droplet model.

\acknowledgements
We thank J. L. Garc{\'\i}a-Palacios, H. Yoshino, and H. Takayama for valuable discussions.
This work was financially supported by the Swedish Natural Science
Research Council (NFR), now assimilated in the Swedish Research Council (VR).

\end{document}